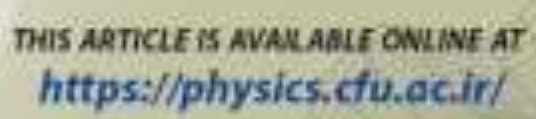

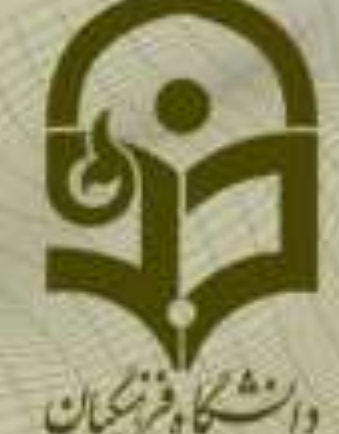

# On the Relativistic Origin of spin: A Case for the "Rest Angular Momentum"

**Habibollah Razmi**[1], **Abbas Mohammad Kazemi**

[1]*Department of Physics, University of Qom, Qom, I. R. Iran.*

**ABSTRACT**

The intrinsic angular momentum, or spin, is a cornerstone of modern physics with profound applications from nuclear magnetic resonance to spintronics. While its mathematical structure within quantum theory is well-defined, its fundamental origin is often less emphasized. This paper revisits the genesis of spin by examining its emergence in relativistic wave equations, its role in the Thomas precession, and its formulation for massless photons in electrodynamics. It is argued that these foundational elements collectively demonstrate that spin is inherently a consequence of relativistic spacetime symmetry, and its full manifestation requires the quantum framework. Consequently, the term "rest angular momentum" offers a more conceptually accurate description, highlighting its origin as an intrinsic property manifest even in an object's rest frame, as dictated by the Poincaré group. Spin, as an invariant property of any object, is the angular momentum an elementary particle possesses in its rest frame, where its orbital angular momentum is zero. This perspective aims to bridge the gap between advanced theoretical concepts and pedagogical clarity, emphasizing that relativity is not merely about high-speed phenomena but is fundamental to the structure of matter itself.

.

1 .Corresponding author:
razmi@qom.ac.ir

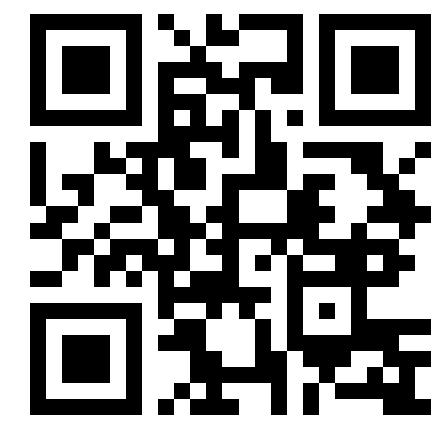

## 1. INTRODUCTION

The discovery of the spin degree of freedom shortly after the formulation of quantum mechanics resolved critical issues in atomic spectroscopy [1]. Today, its implications extend far beyond theoretical significance, underpinning technologies such as NMR, and the burgeoning field of spintronics [2]. Despite its central role, the fundamental physical origin of spin is often obscured in modern treatments, which focus on its mathematical formulation via SU(2) Lie algebra and its postulation in non-relativistic quantum mechanics. The historical difficulty in ascribing a classical "spinning" motion to the electron, which would require a surface velocity exceeding the speed of light, highlighted its non-mechanical, intrinsic nature [3]. While various semi-classical "toy" models have been proposed [4], they remain illustrative rather than fundamentally explanatory to have a realistic classical picture for the spin of fundamental particles. Formally, it is rigorously defined as a form of intrinsic angular momentum and mathematically the spin operator is the generator of intrinsic rotation [5]. With a field-theoretic perspective, Belinfante [6] tried to show that spin is not locally separable from the total angular momentum of the field. Such arguments move decisively away from any classical mechanical picture and instead root the nature of spin in symmetry and conservation laws.

We contend that a coherent understanding of spin's origin is best achieved by tracing its emergence from the framework of special relativity. This work synthesizes evidence from three key areas: (i) the necessity of spin for a consistent relativistic quantum theory, as evidenced by the Dirac equation; (ii) the relativistic kinematical correction of the Thomas precession, which yields the correct electron $g$-factor; and (iii) the derivation of the photon's spin from the relativistic electromagnetic field. This analysis leads us to propose that spin is more aptly termed the *rest angular momentum*, emphasizing its genesis in the irreducible representations of the Poincaré group, which describes the fundamental symmetries of spacetime. This paper aims to clarify that while spin is manifestly a quantum property; its origin is deeply rooted in relativistic symmetry. Our goal is not to claim a purely classical-relativistic origin, but to demonstrate that relativity provides the fundamental bedrock without which spin cannot be understood, a point with significant pedagogical value for physics education.

## 2. Relativistic Wave Equations and the Emergence of Spin

The path to a relativistic quantum mechanics began with the Klein-Gordon equation [7-8]. The Klein-Gordon equation, as a relativistic wave equation for spin-0 particles, proved inadequate as it failed to yield a positive-definite probability density and did not provide a complete description of the electron. Dirac, seeking an equation linear in the first-order time derivative (and hence, in the Hamiltonian), proposed a linear relativistic wave equation in which the wavefunction is a four-component spinor [9]. These extra components (compared to a scalar wavefunction) provide the necessary degrees of freedom to encode a new internal property named as spin. The Dirac equation, formulated to describe the dynamics of relativistic fermions (such as electrons), not only incorporates relativistic effects but also automatically predicts the existence of intrinsic spin-1/2 and its associated properties without any additional postulates. The spin operators and the correct magnetic moment emerged directly from the equation without additional postulates. This stands in stark contrast to non-relativistic quantum mechanics, where spin must be inserted ad hoc into the Schrödinger equation, typically via the Pauli exclusion principle. By focusing on the Lorentz transformation properties of the spinor and computing the commutator algebra of the Hamiltonian with angular momentum operators, the total angular momentum operator separates into two distinct parts: orbital angular momentum and a new internal operator identified as intrinsic spin.

The necessity to the spin can be seen in the example of centrally symmetric potentials for them one expects the angular momentum be conserved. It can be simply shown that the orbital angular momentum operator doesn't commute with the Dirac Hamiltonian unless one considers the spin operator added to the orbital angular momentum operator obeying just the Lie algebra of angular momentum operators and satisfying the covariance under Lorentz transformation. The spin operator emerges directly from the structure of the Lorentz group and the transformation law of the spinor confirming that spin is an intrinsic angular momentum necessitated by relativistic invariance.
Indeed, the Dirac equation predicts spin-1/2 not as an ad hoc postulate but as an inevitable and elegant consequence of unifying quantum mechanics with special relativity. The analysis of the algebraic commutation relations between the Hamiltonian and angular momentum operators provides a rigorous mathematical derivation of this fundamental property.
The fact that a consistent relativistic formulation requires spin demonstrates that its origin is deeply rooted in the union of quantum mechanics and special relativity. The Dirac equation reveals that spin is a property involved with the structure of spacetime itself.

## 3. Thomas Precession and the Relativistic *g*-factor

Further evidence for the relativistic nature of spin comes from the correction of the electron's gyromagnetic ratio. Uhlenbeck and Goudsmit introduction of spin [1] required a *g*-factor of $g_e = 2$ to explain the anomalous Zeeman effect. Within a year, Thomas showed that this factor arises from a purely relativistic kinematical effect: the precession of an accelerating frame relative to an inertial frame, which arises from the non-commutativity of Lorentz boosts, now known as Thomas precession [9]. The Thomas precession frequency exactly halves the naive relativistic correction, resulting in $g_e = 2$. This calculation, derived from special relativity without quantum mechanics, corrects the spin-orbit interaction energy and confirms that the value of the *g*-factor is a relativistic kinematical phenomenon. While quantum electrodynamics provides more precise calculations of the anomalous magnetic moment [10], its foundational origin lies in this relativistic effect.
The Thomas precession, a subtle effect of relativistic kinematics, provides a complementary view, demonstrating how the dynamics of a spinning particle are inherently linked to the structure of spacetime. The precise form of the spin-orbit interaction, confirmed experimentally, stands as a testament to the relativistic origin of spin.

## 4. Spin of the Photon: A Field-Theoretic Perspective

The argument for a relativistic origin is further strengthened by examining the spin of the photon, the quantum of the inherently relativistic electromagnetic field. The total angular momentum $\vec{J}$ of the free electromagnetic field can be decomposed into orbital ( $\vec{L}$ ) and spin ( $\vec{S}$ ) parts [11]:

$$\vec{J} = \frac{1}{4\pi c}\int d^3x \vec{X}\times(\vec{E}\times\vec{B}) = \frac{1}{4\pi c}\int d^3x\left[\vec{E}\times\vec{A}+\sum_{j=1}^{3}E_j(\vec{X}\times\nabla)A_j\right] = \vec{S}+\vec{L} \qquad (1)$$

The first term is an intrinsic position-independent expression; but, the second term depends on the position. The Hermitian quantized (operatorial) form of the first term is:

$$\vec{S} = \frac{1}{8\pi c}\int_{Vol} d^3x\left([\vec{E}\times\vec{A}]-[\vec{A}\times\vec{E}]\right) \qquad (2).$$

Fourier expansion of vector potential operator leads to [7]:

$$\begin{aligned}&\vec{S} = \sum_{\vec{k}}\sum_{m=+1,1}(\hbar m)a^{+}_{\vec{k}m}a_{\vec{k}m}\left(\frac{\vec{k}}{k}\right)\\ &\Rightarrow \vec{S}\left(a^{+}_{\vec{k}m=\pm1}|0\rangle\right) = (\pm 1\hbar)\left(\frac{\vec{k}}{k}\right)(a^{+}_{\vec{k}m=\pm1}|0\rangle)\end{aligned} \qquad (3).$$

Fourier coefficients are creation and annihilation operators and have no $m=0$ component. The photon spin vector has projections along the direction of the $\vec{k}$ vector of $+1\hbar$ (right circularly polarization) or of $-1\hbar$ (left circularly polarization).

Although the spin number of photon is $s=1$, there aren't $2s+1=3$ states available; this is due to the fact that photon has no rest mass and no rest frame.

Once again and based on the above argument on the electromagnetic field which is a purely relativistic field whose quantum particle is photon, it is deduced that the spin which is an intrinsic (position-independent) kind of angular momentum has relativistic origin.

This derivation from the quantized electromagnetic field shows that spin is an intrinsic angular momentum carried by a relativistic field's quanta, not a purely quantum mechanical property.

## 5. The Poincaré Group and the Spin-Statistics Connection

The most profound foundation for this view is the representation theory of the Poincaré group, the group of spacetime symmetries in special relativity. The irreducible representations of this group are classified by two Casimir invariants: mass and spin [14-15]. Elementary particles are defined by these representations, meaning spin is an intrinsic label bestowed by spacetime symmetry, not only by quantum mechanics.

This perspective is reinforced by the spin-statistics theorem [16], which connects a particle's spin to its quantum statistics. While a quantum field theory result, its proof relies critically on the axiom of microcausality; that fields at spacelike separated points must commute (for bosons) or anticommute (for fermions). Since microcausality is a relativistic requirement (that no signal can travel faster than light), the very link between spin and statistics is underwritten by relativity.

## 6. CONCLUSION

The evidence from relativistic quantum mechanics, classical electrodynamics, quantum field theory, and group theory presents a consistent narrative: spin angular momentum is a necessary consequence of incorporating special relativity into physics. It is not a purely quantum mechanical concept but a relativistic one that quantum theory inherits and manifests.

The term "spin" is a historical artifact of a failed classical model. The term "rest angular momentum" is a more physically accurate and pedagogically superior designation. This name emphasizes its defining characteristic: it is the angular momentum an elementary particle possesses in its rest frame, where its orbital angular momentum is zero. This is a direct consequence of its transformation properties under the Lorentz group, the fundamental symmetry of spacetime. Adopting this terminology would clarify its origin and place it within its proper foundational context.

From an educational standpoint, this narrative offers a crucial insight: relativity is not confined to exotic phenomena involving high speeds or strong gravitation, but is fundamentally woven into the fabric of matter itself. Properties like spin, which underpin the periodic table, magnetism, and the very stability of matter, have relativistic roots. A well-known pedagogical example is the golden color of gold, an effect requiring relativistic corrections to quantum mechanics for its explanation. Spin is a similar, yet more fundamental, phenomenon. Emphasizing this connection in teaching can help students appreciate relativity not as an abstract theory for extreme conditions, but as a foundational pillar explaining the everyday world. By clarifying the relativistic origin of spin and its pedagogical significance, this article aims to contribute to bridging the gap between advanced theoretical concepts and accessible educational materials.